\documentclass[twocolumn,showpacs,jcp,superscriptaddress]{revtex4}
\usepackage{graphicx}
\usepackage[usenames]{color}
\usepackage{amssymb}

\usepackage{amsmath}
\usepackage{amsfonts}
\usepackage{mathrsfs}

\renewcommand{\epsilon}{\varepsilon}
\newcommand{\figurewidth}{0.45\textwidth}
\newcommand{\narrowfigurewidth}{0.30\textwidth}

\begin{document}
\title{Dynamics of polymer translocation into a circular nanocontainer through a nanopore}

\author{Kehong Zhang}
%\altaffiliation[]{Author to whom the correspondence should be addressed}
%\email{zhangkh@mail.ustc.edu.cn}
\affiliation{CAS Key Laboratory of Soft Matter Chemistry, Department of Polymer
Science and Engineering, University of Science and Technology of China, Hefei, Anhui
Province 230026, P. R. China}

\affiliation{College of Light-Textile Engineering and Art, Anhui Agriculture University, Hefei, Anhui
Province 230036, P. R. China}

\author{Kaifu Luo}
\altaffiliation[]{Author to whom the correspondence should be addressed}
\email{kluo@ustc.edu.cn}
\affiliation{CAS Key Laboratory of Soft Matter Chemistry, Department of Polymer
Science and Engineering, University of Science and Technology of China, Hefei, Anhui
Province 230026, P. R. China}

\date{\today}
%%%%%%%%%%%%%%%%%%%%%%%%%%%%%%%%%%%%%%%%%%%%%%%%%%%%%%%%%%%%%%

\begin{abstract}

Using Langevin dynamics simulations, we investigate the dynamics of polymer translocation into a circular nanocontainer through a nanopore under a driving force $F$. We observe that the translocation probability initially increases and then saturates with increasing $F$, independent of $\phi$, which is the average density of the whole chain in the nanocontainer. The translocation time distribution undergoes a transition from a Gaussian distribution to an asymmetric distribution with increasing $\phi$. Moreover, we find a nonuniversal scaling exponent of the translocation time as chain length, depending on $\phi$ and $F$. These results are interpreted by the conformation of the translocated chain in the nanocontainer and the time of an individual segment passing through the pore during translocation.

\end{abstract}

%%%%%%%%%%%%%%%%%%%%%%%%%%%%%%%%%%%%%%%%%%%%%%%%%%%%%%%%%%%%%%

\maketitle

\section{INTRODUCTION}

Biopolymer translocation across a membrane through a nanopore has attracted broad interest because it is essential to many biological processes, such as DNA and RNA translocation across nuclear pores, protein transport through membrane channels, virus infection \cite{Alberts}. In a seminal paper, it has been experimentally demonstrated that an external applied electric field can effectively drive a single-stranded DNA and RNA molecules through the $\alpha$-hemolysin channel and that the passage of each molecule is signaled by a blockade in the channel current \cite{Kasianowicz96}.
Due to its numerous revolutionary applications, including rapid DNA sequencing, filtration, gene therapy, flow-injection problems, and controlled drug delivery \cite{Meller03,Branton}, considerable experimental \cite{Meller03,Meller2,Meller3,Akeson,Wanunu,Sauer-Budge,Kasianowicz2,Kasianowicz3,Henrickson,Robertson,Li,Fologea,
Storm1,Storm2}, theoretical and numerical \cite{Sung,Muthukumar1,Chuang,Kantor,Huopaniemi,Luo1,Luo2,Dubbeldam1,Dubbeldam2,Vocks,Luo10,Luo3,Sakaue1,Gauthier,
Bhattacharya,Chen,Sun,Luo4,Muthukumar2,Kong,Muthukumar3,Sakaue3,Luo5,Luo6} studies have been devoted to polymer translocation.

Based on both the basic physics as well as a technology design perspective, the scaling of the average translocation time $\tau$ with the polymer length $N$, $\tau \sim N^\alpha$, is an important measure in the study of polymer translocation. The scaling exponent $\alpha$ reflects the translocation efficiency. Standard equilibrium Kramers analysis \cite{Kramers} of diffusion through an entropic barrier discovers $\tau \sim N^2$ for unbiased translocation and  $\tau \sim N$ for driven translocation (assuming friction is independent of $N$) \cite{Sung,Muthukumar1}. However, the scaling behavior $\tau \sim N^2$ is incorrect at least for a self-avoiding polymer. The reason is that the translocation time is shorter than the Rouse equilibration time of a self-avoiding polymer, $\tau_R \sim N^{1+2\nu}$, where the Flory exponent $\nu = 0.588$ in three dimensions (3D) and 0.75 in two dimensions (2D) \cite{deGennes,Rubinstein}. In contrast, the scaling behavior $\tau_R \sim N^{1+2\nu}$ for both phantom and self-avoiding polymers was demonstrated by Chuang \textit{et al.} \cite{Chuang} and Luo \textit{et al.} \cite{Luo1} based on  a 2D lattice model, the Fluctuating Bond and Langevin Dynamics (LD) \cite{Huopaniemi} models with the bead-spring approach, respectively. For driven translocation by a transmembrane electric field, a lower bound $\tau \sim N^{1+\nu}$ for the translocation time was provided by Kantor and Kardar \cite{Kantor}. Recent results \cite{Luo3,Luo10} further show that there is a crossover from $\tau \sim N^{1.37}$ for faster translocation processes to $\tau \sim N^{1+\nu}$ for slower translocation in 3D.

In addition to a transmembrane electric field, the driving force can also be provided by a pulling force exerted on the end of a polymer \cite{Kantor,Huopaniemi2}, binding particles (chaperones) \cite{Ralf,Yu}, longitudinal drag flow in a  nanochannel \cite{Luo11}, or geometrical confinement of the polymer \cite{Park,Muthukumar4,Cacciuto2,Gopinathan,Luo7}.

For driven translocation by a transmembrane electric field, however, above physical pictures are based on translocation into an unconfined \textit{trans} side. Very little attention is paid to the dynamics of translocation into confined environments \cite{Luo4}. Even for the case of an unconfined \textit{trans} side, the translocated chain is highly compressed during the translocation process because it does not have enough time to diffusion away from the exit of the pore under fast translocation conditions \cite{Luo1,Luo3}, and thus even more severe nonequilibrium effects are expected under confinement.

Polymer translocation into a unclosed confined environment, such as a slit or a fluidic channel \cite{Luo4}, is subject to a large entropic penalty due to reduced accessible degrees of freedom, which should dramatically affect the translocation dynamics. In addition, studies on translocation into confined geometries will shed light on the dynamics of the packaging of DNA inside virus capsid \cite{Smith,Gennes,Arsuaga}, where the entropic penalty is from both the limited volume and the additional self-exclusion of the chain. Using the stochastic rotation dynamics (also called multiparticle collision dynamics) simulations \cite{Kapral}, Ali \textit{et al.} \cite{Ali} have investigated the packaging of flexible and semiflexible polymers. However, they only focused on the packing rate and the pauses in the packaging process. Many important details are still not clear, such as (a) the translocation probability, (b) the distribution of the translocation time, and (c) the average translocation time as a function of the density of the chain, the chain length, and the driving force.
To this end, we investigate the dynamics of polymer translocation into 2D circular nanocontainer using Langevin dynamics simulations.

The paper is organized as follows. In Sec. II, we briefly describe our model and the simulation technique. In Sec. III, we present our results. Finally, the conclusions and discussion are presented in Sec. IV.

\section{MODEL AND METHODS}

In our simulation, we model the polymer chains as bead-spring chains of Lennard-Jones (LJ) particles with the Finite Extension Nonlinear Elastic (FENE) potential \cite{Kremer}. Excluded volume interactions between beads is modeled by a short range repulsive LJ potential:
\begin{equation}
        U_{LJ}=4\varepsilon[(\frac{\sigma}{r})^{12} - (\frac{\sigma}{r})^6] + \varepsilon,
\end{equation}
for $r \leq 2^{1/6}\sigma$ and 0 for $r > 2^{1/6}\sigma$. Here, $\sigma$ is the diameter of a monomer and $\varepsilon$ is the potential depth. The connectivity between neighboring beads is modeled as a FENE spring with
\begin{equation}
     U_{FENE}(r)=-\frac{1}{2}kR_0^2\ln(1-r^2/R_0^2),
\end{equation}
where $r$ is the distance between consecutive monomers, $k$ is the spring constant and $R_0$ is the maximum allowed separation between connected monomers.

\begin{figure}
\includegraphics*[width=\figurewidth]{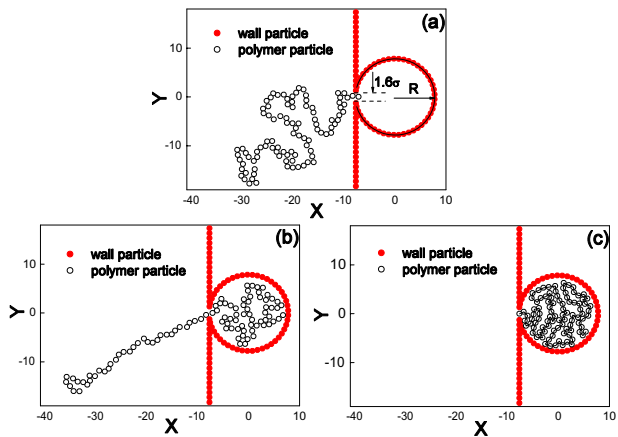}
\caption{Schematic representation of polymer translocation through a pore into a 2D spherical compartment with radius $R$ through a nanopore driven by an external electric force $F=5$: (a) before translocation, (b) during translocation, and (c) after translocation.
The width of the pore is $w=1.6\sigma$.
        }
\label{Fig1}
\end{figure}

As shown in Fig. \ref{Fig1}, we consider a two-dimensional (2D) geometry, where the chain is translocating into a nanocontainer through a nanopore driven by a transmembrane electric field. The circular nanocontainer with a pore of width $w=1.6\sigma$ is formed by stationary wall particles within a distance $\sigma$ from each other. Between all monomer-wall particle pairs, there exits the same short range repulsive LJ interaction as depicted above. In Langevin dynamics simulation, each monomer is subjected to conservative, frictional and random forces, respectively.
\begin{equation}
        m{\bf \ddot{r}}_i=-{\bf\nabla}({U}_{LJ}+{U}_{FENE})+{\bf F}_{\textrm{ext}}-\xi {\bf v}_i+{\bf F}_i^R,
\end{equation}
where $m$ is the monomer's mass, $\xi$ is the friction coefficient, ${\bf v}_i$ is the bead's velocity, and ${\bf  F}_i^R$ is the random
force satisfying the fluctuation-dissipation theorem. The external force is expressed as ${\bf F}_{ext} = F\hat{x}$, where $F$ is the external force strength exerted exclusively on the monomer in the pore, and $\hat{x}$ is a unit vector in the direction along the pore axis. This driving force mimics the role of the portal molecular motor for packaging of DNA molecules into capsid.

In the present work, we use the LJ parameters $\varepsilon$, $\sigma$, and $m$ to fit the system energy, length and mass scales, respectively. As a result, the corresponding time scale and force scale are given by $t_{LJ}=(m\sigma^2/\epsilon)^{1/2}$ and $\epsilon/\sigma$ respectively, which are order of ps and pN, respectively.
We set $k_{B}T=1.2\epsilon$, which means the interaction strength $\epsilon$ to be $3.39 \times 10^{-21}$ J at actual temperature 295 K. This leads to a force scale of 3.3 pN for a chain with segmental length $\sigma =1$ nm.
In our simulation, the dimensionless parameters are chosen to be $R_0=2$, $k=7$, $\xi=0.7$.
Then, the Langevin equation is integrated in time by the method proposed by Ermak and Buckholz \cite{Ermak}.

Initially, the first monomer is placed at the pore center, as shown in Fig. \ref{Fig1}(a), while the remaining monomers are undergoing thermal collisions described by the Langevin thermostat to obtain an equilibrium configuration. The translocation time $\tau$ is defined as the time duration between the beginning of the translocation and the last monomer of the chain entering into the circular nanocontainer, see Fig. \ref{Fig1}(c). Typically, the averaging is done over 2000 successful translocation events.

\section{RESULTS AND DISCUSSIONS}

\subsection{Theory}

The conformation of a linear chain confined in a space of characteristic size lower than its radius of gyration has been broadly studied \cite{Daoud,Cassasa,Cacciuto1,Sakaue4,Micheletti,Grosberg}. The scaling behavior of free energy for polymer in full three-dimensional confinement, such as a spherical cavity, is different compared with slit-like confinement and tube-like confinement \cite{Cacciuto1,Sakaue4,Grosberg}.

Following Refs. \onlinecite{Sakaue4} and \onlinecite{Grosberg}, we first consider the equilibrium behavior of a linear chain of length $N$ confined into a two-dimensional circular nanocontainer of the radius $R\ll R_g$.
Here, $R_g \sim N^{\nu_{2D}}\sigma$ is the radius of gyration of the chain without confinement, with $\nu_{2D}=0.75$ being the Flory exponent in 2D \cite{deGennes,Rubinstein}. The notation `$\sim$' means that two quantities are proportional to each other.
The density of the bead in the nanocontainer is
\begin{equation}
 \phi={N\sigma^2}/(2R)^2.
\label{eq1}
\end{equation}
According to the blob picture, the chain will form blobs with size $\xi_b$, and each blob contains $g \sim (\xi_b/\sigma)^{1/{\nu_{2D}}}$ monomers. The density of the monomer in a blob is
\begin{equation}
 \phi_{blob}=g\sigma^2/\xi_b^2,
\label{eq2}
\end{equation}
which should be the same as that of the bead in the whole nanocontainer, $\phi_{blob}=\phi$.  Therefore, based on Eqs. (\ref{eq1}) and (\ref{eq2}), the blob size is
\begin{equation}
\xi_b \sim \sigma \phi^{-\nu_{2D}/(2\nu_{2D}-1)}=\sigma \phi^{-1.5},
\label{eq3}
\end{equation}
which indicates that the blob size decreases with increasing $\phi$.

Then, the number of blobs is
\begin{equation}
n_b = N/g \sim N\phi^{1/(2\nu_{2D}-1)}=N\phi^2.
\label{eq4}
\end{equation}
The free energy cost is proportional to the number of blobs, and thus in units of $k_BT$ it is %$\mathcal{F} = N\phi^{1/(2\nu_{2D}-1)}$.
\begin{equation}
\mathcal{F} \sim N\phi^{1/(2\nu_{2D}-1)}=N\phi^2.
\label{eq5}
\end{equation}
This shows that the free energy of the chain increases with $\phi^2$.

For polymer translocation into a circular nanocontainer, an external driving force $F$ is necessary to overcome the entropic force from the already translocated beads at \textit{cis} side. With increasing the time, the density of the chain in the nanocontainer increases. Once the translocated portion of the chain feels to be confined, it will form blobs and the blob size becomes smaller and smaller with the time. Thus, the entropic repulsion $f(\phi(t))$ exerted by the already translocated monomers increases with the time during the translocation process.
Owing to the highly nonequilibrium nature of the translocation process, it is difficult to calculate $f(\phi(t))$.

To examine the the effect of the confinement on the translocation time, we define $\phi$ as the chain density in the circular nanocontainer after the translocation and $f(\phi)$ as the average resisting force for the whole translocation process.
Then, we can write the translocation time as
\begin{equation}
\tau \sim N^\alpha/[F-f(\phi)]=N^\alpha/[F(1-f(\phi)/F)],
\label{eq6}
\end{equation}
where $F$ is the external driving force in the pore and $\alpha$ is the scaling exponent of $\tau$ with chain length $N$. Here, $F(1-f(\phi)/F)$ is the effective driving force. For an unconfined translocation ($R \rightarrow \infty $), translocation time is
\begin{equation}
\tau_\infty \sim N^\alpha/F,
\label{eq8}
\end{equation}
and then we obtain
\begin{equation}
1-\frac{\tau_\infty}{\tau} \sim f(\phi)/F.
\label{eq9}
\end{equation}
Based on this relationship, we can explore the dependence of the average resisting force $f(\phi)$ on the density of the chain $\phi$.

\subsection{Numerical results}

During the translocation process, the effective driving force is $F(1-f(\phi)/F)$. In our simulations, if $F<1$, the translocation probability is too small and the translocation time is too long, particularity for higher $\phi$. To obtain effective statistical average for translocation probability and translocation time, we present results for $F\ge 1$.

\subsubsection{Translocation probability as a function of the driving force $F$ and the density of the chain $\phi$}

\begin{figure}
\includegraphics*[width=\figurewidth]{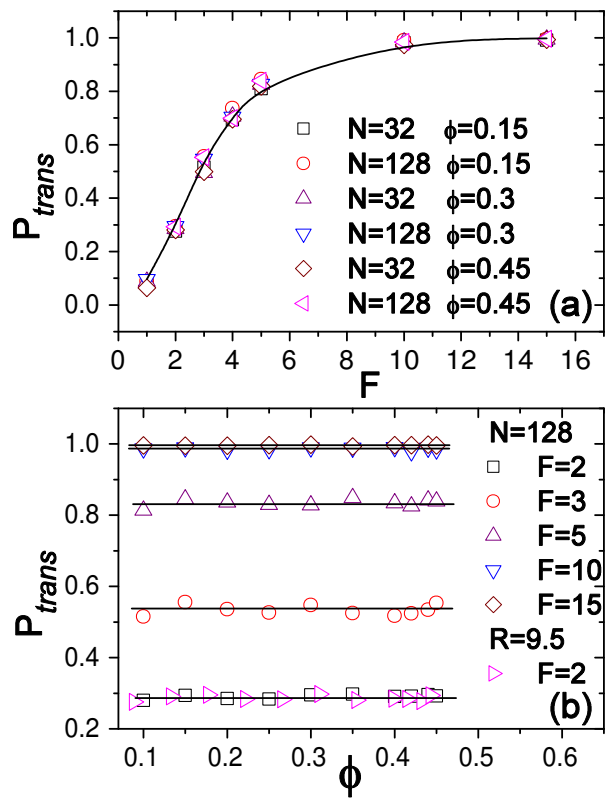}
\caption{(a) Translocation probability as function of the driving force $F$ for $N=32,128$ and different density of the chain $\phi$. (b) Translocation probability as function of the density of the chain $\phi$ for chain length $N=128$ ($F=2$, 3, 5, 10 and 15) and $R=9.5$ ($F=2$).
        }
\label{Fig2}
\end{figure}

We refer to a successful translocation as the event when the chain fully enters into the circular nanocontainer, as shown in Fig. \ref{Fig1}(c). Otherwise, the translocation is considered as unsuccessful.
Fig. \ref{Fig2}(a) depicts the translocation probability, denoted as $P_{trans}$, as a function of the driving force $F$ for different chain lengths ($N=32$ and $128$) and density of the chains ($\phi=0.15$, $0.3$ and $0.45$ through the change of $R$). With increasing the driving force $F$, $P_{trans}$ increases rapidly first, and then slowly approaches saturation at larger $F$, almost independent on $N$ and $\phi$.
As shown in Fig. \ref{Fig2}(b), we also observe that for a fixed $N$ value, $P_{trans}$ is independent of $\phi$ we examined for different driving forces.

If the radius of the circular nanocontainer $R$ is fixed, $\phi$ changes with $N$. Intuitively, increasing the density of the chain will lead to the decrease of $P_{trans}$ due to the increase of the resisting force. However, we find that $P_{trans}$ is also independent of $\phi$ for $R=9.5$ and $F=2$.
During the translocation process, the density of the chain $\phi$ of translocated monomers in the nanocontainer increases with the time. But the effect of $\phi$ at the early time of the translocation process is negligible, which dominates $P_{trans}$.

\begin{figure}
\includegraphics*[width=\figurewidth]{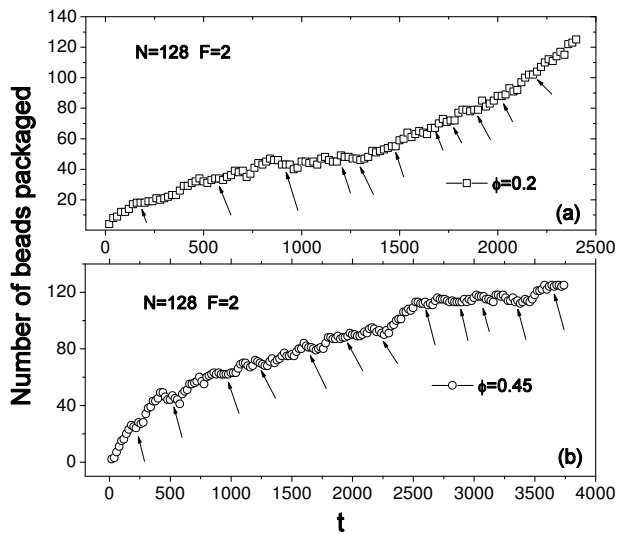}
\caption{Number of beads packaged against time for the chain length $N=128$, the driving force $F=2$, and different $\phi$. Here, all the data points are not averaged, and are from a typical successful translocation event. The pause is defined as the situation that the number of packaged beads is constant with time, and the typical pause is marked by arrow in the figures. The average time duration of pauses for $\phi=0.45$ is longer than those for $\phi=0.2$.
        }
\label{Fig3}
\end{figure}

Even if $\phi$ is large, which means the resisting force is strong at the late time of the translocation process, the chain can experience backward motion to decrease the resisting force. Thus, it is still difficult for the first monomer to exit the pore, and $P_{trans}$ is independent of $\phi$. This is can be seen in Fig. \ref{Fig3}, where a typical plot of the number of beads packaged against time for a successful translocation event.  For the chain length $N=128$ and the driving force $F=2$, we observe pauses for both $\phi=0.2$ and 0.45, where the number of packaged beads is constant with time. The average time duration of pauses for $\phi=0.45$ is longer than those for $\phi=0.2$. We should point out that pauses are also observed in the stochastic rotation dynamics simulations \cite{Ali}, as well as in the experiments on DNA packaging \cite{Smith}.

\subsubsection{Distribution of the translocation time}

\begin{figure}
\includegraphics*[width=\figurewidth]{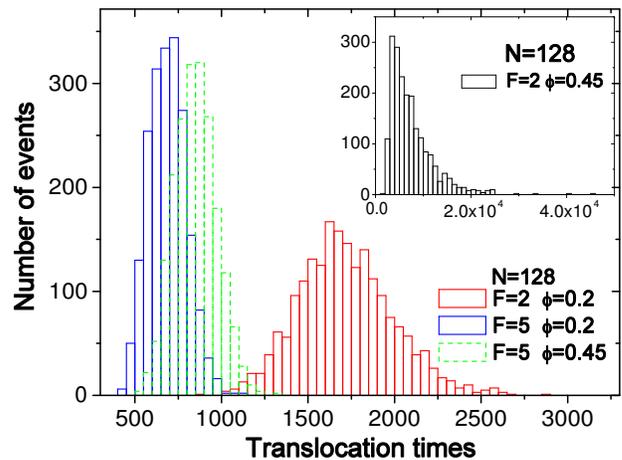}
\caption{Histogram of translocation times for the chain length $N=128$ and different $F$ and $\phi$. Here,  $F = 2$ and $5$ and $\phi = 0.2$ and 0.45. The inset shows the distribution of translocation times for $F = 2$ and $\phi = 0.45$.
 }
\label{Fig4}
\end{figure}

Fig. \ref{Fig4} shows the distribution of translocation times for $N=128$ and different $F$ and $\phi$.
Under the driving force $F=5$, the histograms for both $\phi=0.2$ and 0.45 show narrow Gaussian distributions, but the position of the peak moves to a longer time for $\phi=0.45$ compared with that for $\phi=0.2$. With decreasing the driving force to $F=2$, for $\phi=0.2$ the histogram still looks like Gaussian distribution but is wider by comparison with the cases for $F=5$. However, for $\phi=0.45$ the distribution is much wider and highly asymmetric, see the inset of Fig. \ref{Fig4}. These distributions can be understood by taking into account the effective driving force $F(1-f(\phi)/F)$. With increasing $\phi$, the effective driving force $F(1-f(\phi)/F)$ decreases, which leads to the peaks of the distributions moving to longer times. For $F=2$ and $\phi=0.45$, the effective driving force becomes quite small during the late time of the translocation process, resulting in an highly asymmetric distribution.

\subsubsection{Translocation time as a function of the density of the chain $\phi$}

\begin{figure}
\includegraphics*[width=\figurewidth]{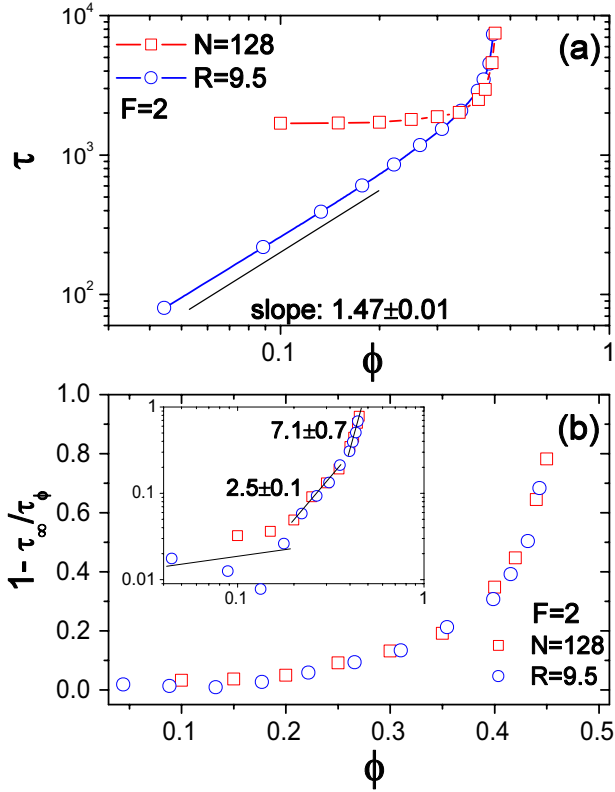}
\caption{(a) Translocation time $\tau$ as a function of the density of the chain, $\phi$, under driving force $F=2$ for a fixed chain length $128$ or a fixed radius of the nanocontainer $R=9.5$. (b) $1 - (\tau_\infty/\tau)$ as a function of $\phi$.
        }
\label{Fig5}
\end{figure}

Fig. \ref{Fig5}(a) shows the translocation time $\tau$ as a function of the density of the chain $\phi$ under the driving force $F=2$ for a fixed chain length $N=128$ or a fixed radius of the nanocontainer $R=9.5$. For a fixed chain length $N=128$, we increase $\phi$ by decreasing $R$. For this case, we find that $\tau$ initially increases very slowly with $\phi$, and it increases very rapidly after a critical density of the chain, $\phi_c\approx 0.4$. For a fixed $R=9.5$, $\phi$ is increased by increasing $N$. For $\phi < 0.2$, we find $\tau \sim \phi^{1.47}$, implying $\tau \sim N^{1.47}$. The scaling exponent is close to $2\nu$ for translocation into unconfined system, reflecting that $\phi$ does not affect the scaling exponent for $\phi < 0.2$ due to $F(1-f(\phi)/F)\approx F$. However, $\tau$ rapidly increases with $\phi$ particularly after $\phi_c$  because $f(\phi)$ becomes important.

Fig. \ref{Fig5}(b) shows $1-(\tau_\infty/\tau)$ as a function of $\phi$ for $F=2$. Here, $\phi$ varies by either changing $R$ or $N$. The data points for both cases almost collapse on one curve. With increasing $\phi$, $1-(\tau_\infty/\tau)$, which is proportional to $f(\phi)/F$, increases slowly at first, followed by a rapid increases. For strong confinement $\phi=0.45$, the average resisting force $f(\phi)$ is about 80\% of the driving force $F$. This indicates that the effective driving force $F(1-f(\phi)/F)$ is quite small ($2\times20\%=0.4$), leading to asymmetric distribution with long tail of translocation times shown in the inset of Fig. \ref{Fig4}. From the inset of Fig. \ref{Fig5}(b), $f(\phi)/F \sim \phi^{\beta}$, where $\beta$ is very small for the weak confinement ($\phi<0.2$), 2.5 for the moderate confinement ($0.2<\phi<0.4$) and 7.1 for the strong confinement ($\phi>0.4$), reflecting the very rapid increase of the resisting force with $\phi$ for larger $\phi$.

%For translocation process closed to equilibrium $f(\phi) \sim \phi^{1/(2\nu_{2D}-1)} \sim \phi^2$. The numerical exponents do not agree %with this result, which shows the translocation into sphere is a highly non-equilibrium process.

%\begin{figure}
%\includegraphics*[width=\figurewidth]{Fig5b.eps}
%\caption{(Color online) For a typical translocation process, the resisting force felt by the bead in the pore just after the translocating %into the cavity as a function of the number of the packed beads for $N=128$, $\phi=0.4$ and different driving forces.
%        }
%\label{Fig5b}
%\end{figure}

\subsubsection{Translocation time as a function of the chain length $N$}

\begin{figure}
\includegraphics*[width=\figurewidth]{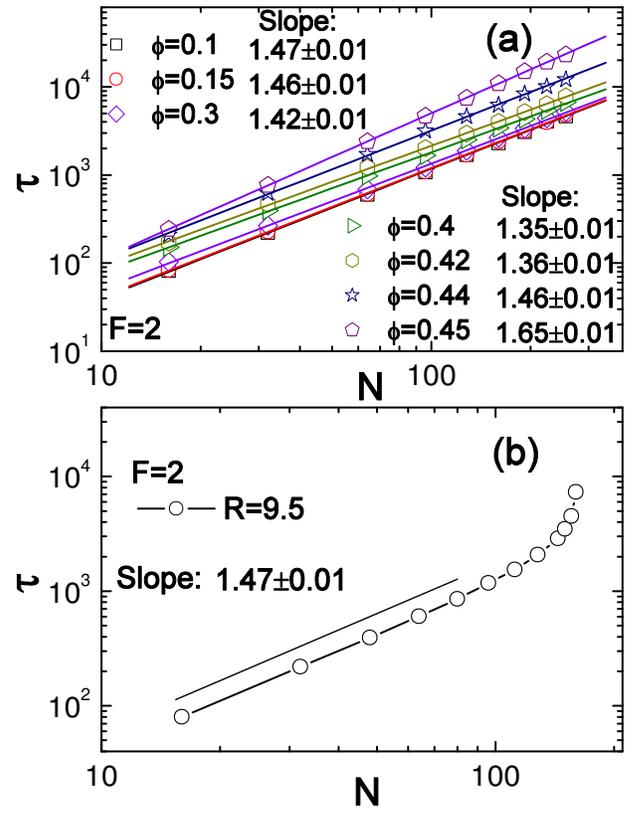}
\caption{Translocation time $\tau$ as a function of the chain length $N$ under the driving force $F=2$: (a) for different density of the chain, $\phi$, (by changing $R$), and (b) for $R=9.5$.
        }
\label{Fig6}
\end{figure}

\begin{figure}
\includegraphics*[width=\figurewidth]{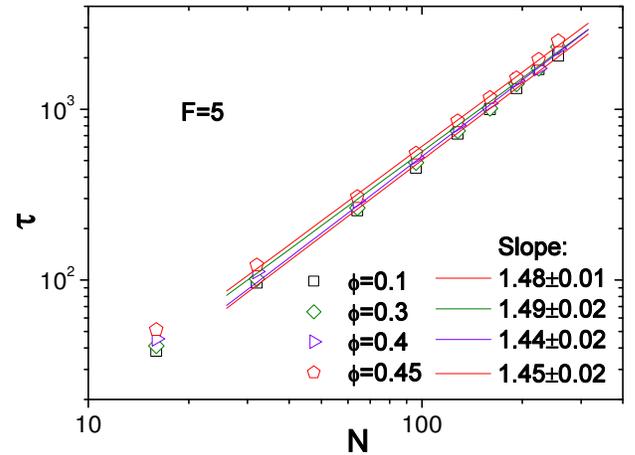}
\caption{Translocation time $\tau$ as a function of the chain length $N$ for different density of the chain $\phi$ and different chain lengths $N$ under the driving force $F=5$.
        }
\label{Fig7}
\end{figure}

Fig. \ref{Fig6}(a) shows $\tau$ as a function of the chain length $N$ under the driving force $F=2$ for different density of the chain $\phi$. In the simulations, for a fixed $\phi$ we changed $R$ for different $N$. We find $\tau \sim N^\alpha$ with $\alpha$ being the scaling exponent. %
For a weak confinement, $\phi = 0.1$ we observe $\alpha=1.47$, in agreement with the previous study for translocation into an unconfined system \cite{Huopaniemi}, where $\alpha=2\nu=1.5$ for $N \leq 200$ in 2D. This result also indicates that the weak confinement does not change $\alpha$ due to the effective driving force $F(1-f(\phi)/F)\approx F$. However, with increasing $\phi$, the dynamics is different for translocation into a nanocontainer. We observe that $\alpha$ slightly decreases first from $\alpha \approx 2\nu$ and then increases with increasing $\phi$, such as $\alpha =1.65$ for $\phi=0.45$. With increasing $\phi$ from 0.40 to 0.45, the resisting force $f(\phi)$ increases rapidly with $\phi$, leading to a great decrease of the effective driving force $F(1-f(\phi)/F)$. This is the reason for the increase of $\alpha$.
Here, the exponent $\alpha$ increases with decreasing the effective driving force, which also confirms previous results on forced translocation inside an unconfined region under different driving forces \cite{Luo3,Edmonds} and results on translocation into two-dimensional fluidic nanochannels for a fixed driving force and different channel widths \cite{Luo4}. Recently, Lehtola \textit{et al.} \cite{Lehtola} also observe a nonuniversal scaling exponent for forced translocation inside an unconfined region, however, the exponent $\alpha$ increases with increasing the driving force.

For a stronger driving force $F=5$, the scaling exponent $\alpha$ almost does not change even for $\phi=0.45$, see Fig. \ref{Fig7}. This is because the driving force is dominant compared with the resisting force during the translocation process, namely, $F(1-f(\phi)/F)\approx F$ is still correct for larger $F$.

As a comparison, Fig. \ref{Fig6}(b) shows $\tau$ as a function of the chain length $N$ under the driving force $F=2$ for a fixed radius of the nanocontainer $R=9.5$, where the resisting force $f(\phi)$ increases slowly with $N$ for short $N$, and thus we still observe $\alpha \approx 2\nu$ for short chains. For longer chains, the resisting force $f(\phi)$ increases rapidly with $N$ due to both the limited volume and the additional self-exclusion of the chain, leading to a much more rapid increase of the translocation time $\tau$.

\subsubsection{Translocation time as a function of the driving force $F$}

In previous study for translocation into the unconfined system \cite{Huopaniemi,Luo3}, it has been shown that $\tau \sim F^{-1}$ for lower forces and $\tau \sim F^{-0.8}$ for higher driving  forces.

\begin{figure}
\includegraphics*[width=\figurewidth]{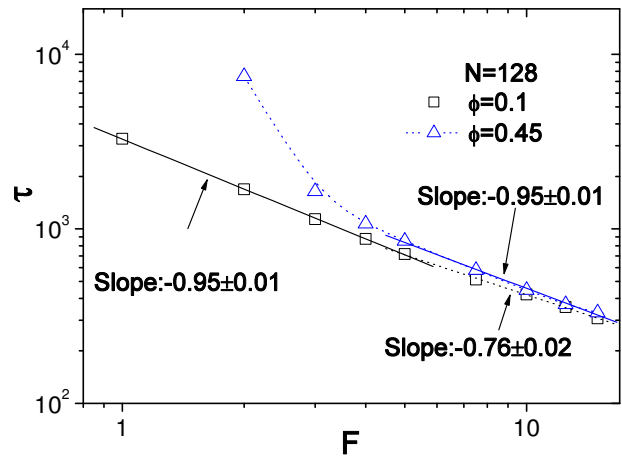}
\caption{Translocation time $\tau$ as a function of the driving force $F$ for chain lengths $N=128$ under the density of the chain $\phi =0.1$ and $0.45$ in 2D.
        }
\label{Fig8}
\end{figure}

For translocation into confined nanocontainer, the chain density in the nanocontainer, $\phi$, increases with time, leading to that the resisting force $f(\phi)$ increases with time for a typical translocation event. In addition, the value of $f(\phi)$ increases with $\phi$ as shown in Fig. \ref{Fig5}(b). For the lower density of the chain such as $\phi=0.1$, the effective driving force $F(1-f(\phi)/F) \approx F$, and thus we observe the similar results, see Fig. \ref{Fig8}, compared with translocation into unconfined systems \cite{Huopaniemi}. For $F \le 5$, we observe $\tau \sim F^{-1}$ before crossing over to $\tau \sim F^{-0.76}$ for $F > 5$. The scaling of $\tau \sim F^{-1}$ can be understood by considering the balance of the frictional damping force with the driving force $F$. For $F > 5$, a different scaling behavior is observed due to a pronounced non-equilibrium effect.
However, for $\phi=0.45$, $\tau$ rapidly decreases with $F$ for $F\le 4$ because $f(\phi)$ is important and cannot be negligible for the translocation dynamics. For $F \ge 5$, $f(\phi)$ becomes less and less important due to $F(1-f(\phi)/F) \approx F$ for larger $F$, and thus $\tau \sim F^{-1}$ is observed.

\begin{figure}
\includegraphics*[width=\figurewidth]{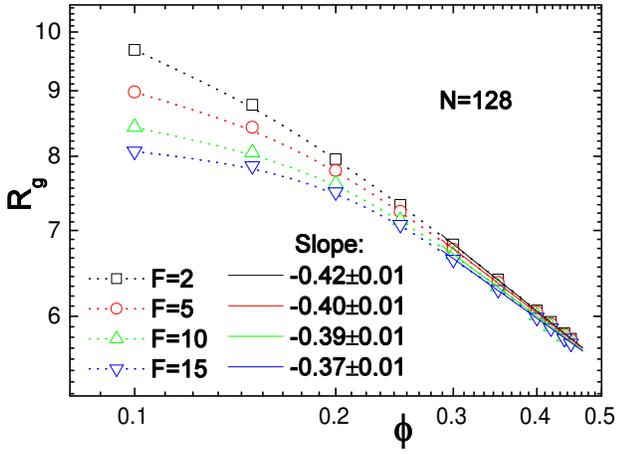}
\caption{The radius of gyration of the chain $R_g$ at the moment just after the translocation for $N = 128$ under different driving F force versus the density of the chain $\phi$.
        }
\label{Fig9}
\end{figure}

\begin{figure}
\includegraphics*[width=\narrowfigurewidth]{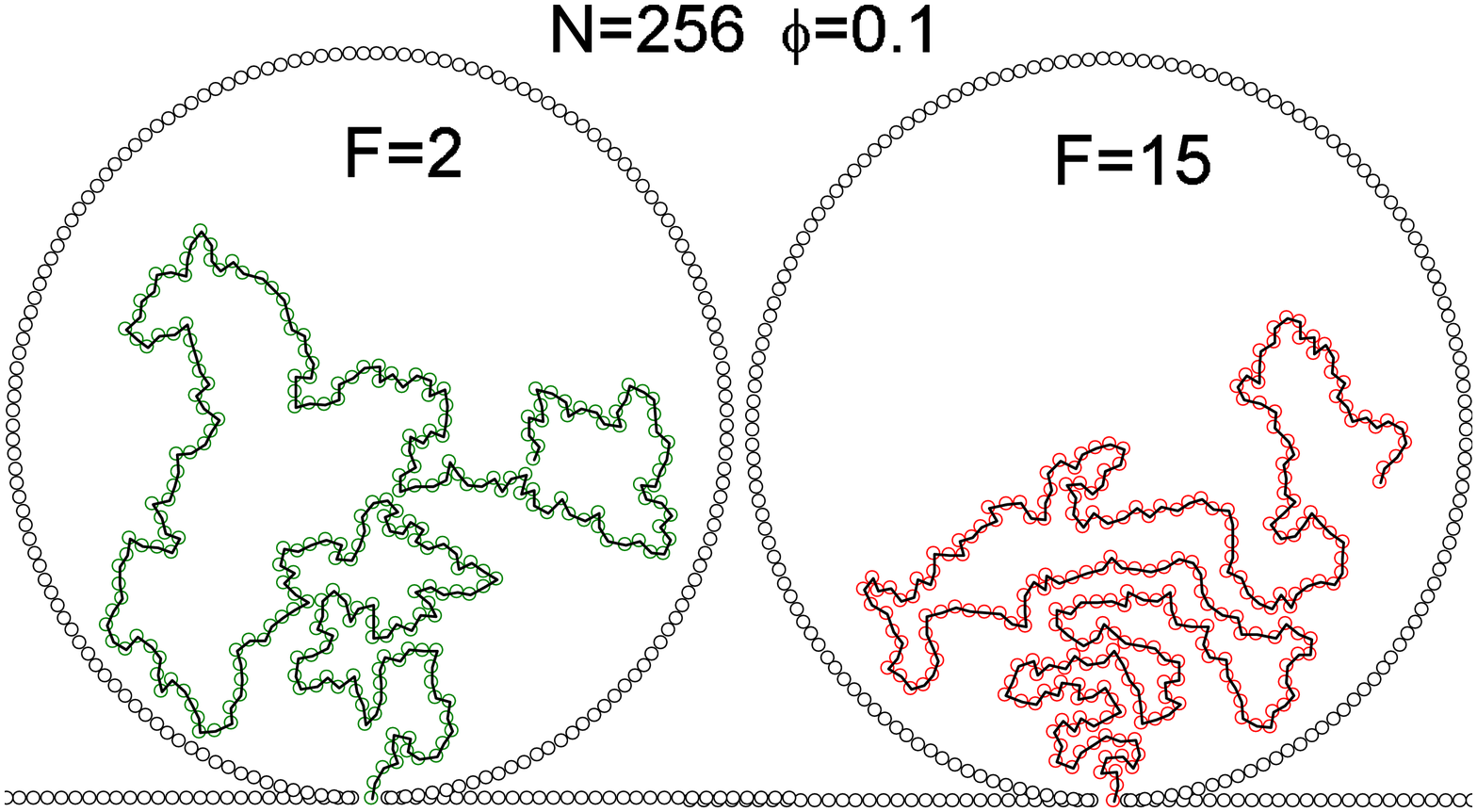}
\includegraphics*[width=\figurewidth]{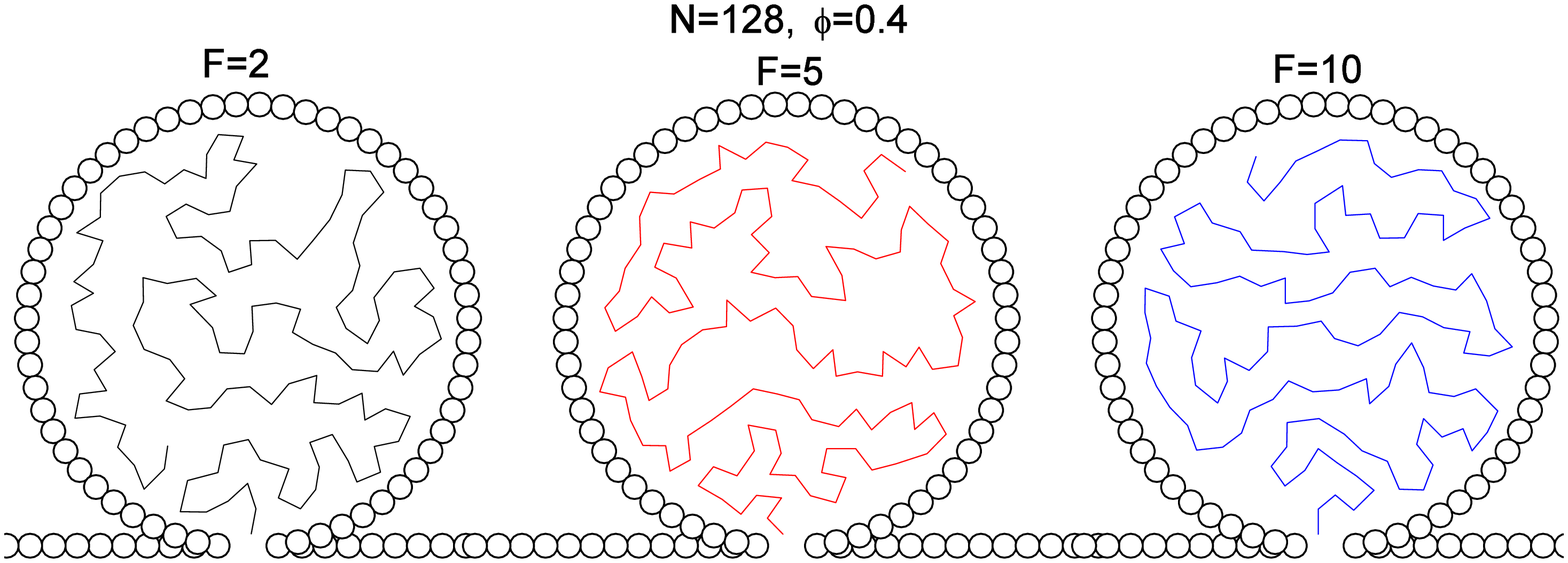}
\caption{The chain conformation at the moment just after the translocation for the chain length $N = 256$ and the density of the chain $\phi = 0.1$ under the driving forces $F = 2$ and $15$ in 2D, respectively.
        }
\label{Fig10}
\end{figure}

Fig. \ref{Fig9} shows the radius of gyration of the chain, $R_g$, at the moment just after the translocation for $N = 128$ as a function of the density of the chain $\phi$ under different driving forces.
For the weak confinement regime with small density of the chain $\phi$, the radius of gyration of the translocated chain $R_g$ under the strong driving force is much smaller than that under the weak driving force. The reason is that the translocated chain are more compressed nearby the entrance for strong driving forces, see the chain conformations for $N=256$ and $\phi=0.1$ under $F=2$ and $F=15$ shown in Fig. \ref{Fig10}.

For equilibrated chains, the radius of gyration of the chain in sphere is proportional to the radius of sphere under the strong confinement, namely, $R_g \sim (N/\phi)^{1/2} \sim N^{1/2}\phi^{-1/2}$.
The numerical exponent of $R_g$ with $\phi$ increases from -0.42 for $F = 2$ to -0.37 for $F = 15$. Due to the non-equilibrium effect, the chain is not homogenous distributed in the nanocontainer, and thus the simulation numerical exponent does not agree with -1/2. Particularly, for larger $F$, the segments of the chain are locally oriented along the direction perpendicular to the axis of the pore.

\subsubsection{Waiting time distributions}

It is necessary to explore the dynamics of a single segment passing through the pore into the nanocontainer. The time of an individual segment passing through the pore is significantly affected by the non-equilibrium nature of the translocation process. We numerically calculated the waiting times for each monomer of the chain. The waiting time of bead $s$ is defined as the average time between the events that bead $s$ and bead $s + 1$ first pass through the exit of the pore.

\begin{figure}
\includegraphics*[width=\figurewidth]{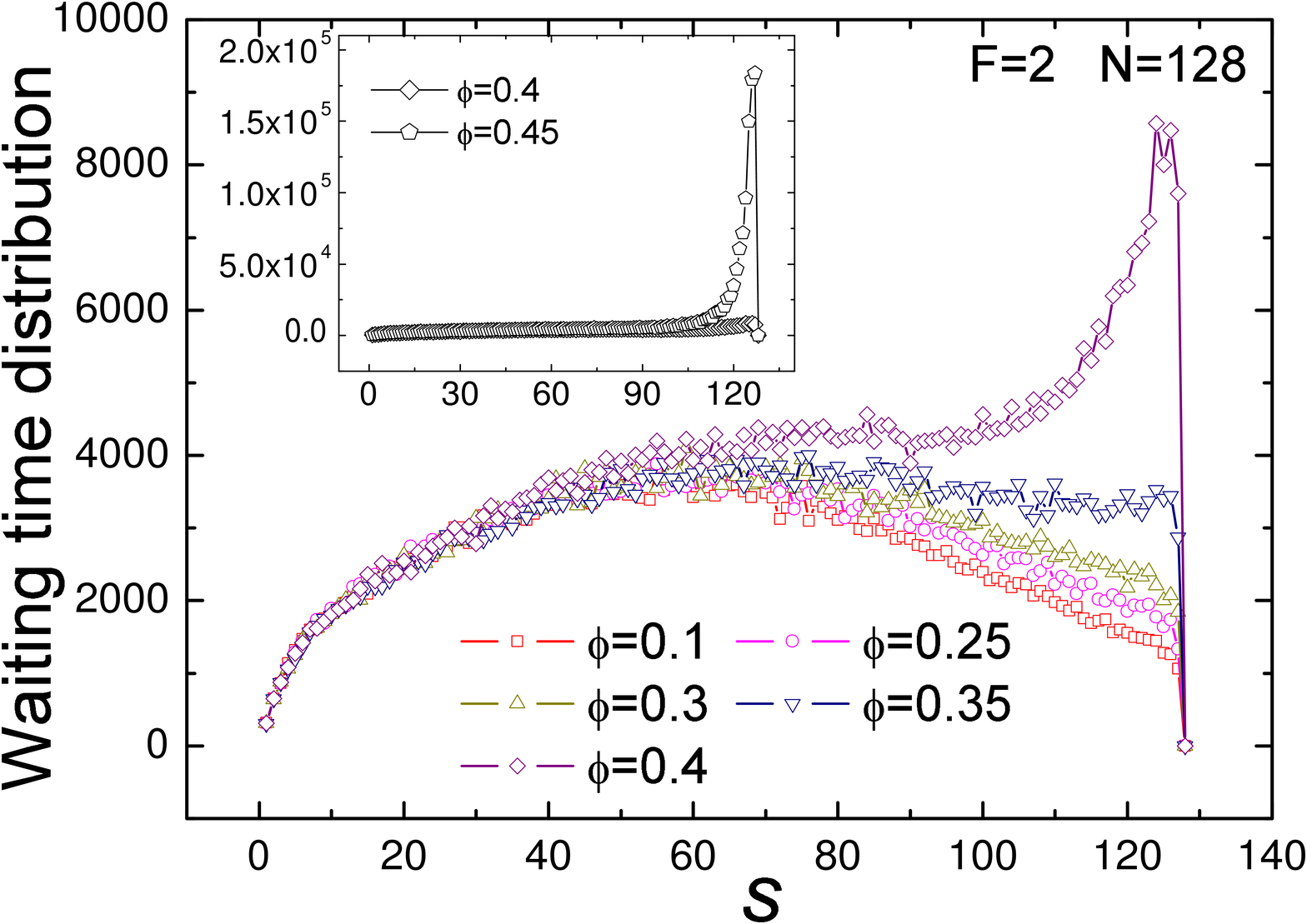}
\caption{Waiting time distribution for different densities of the chain $\phi$. Here,  the chain length is $N=128$, and the driving force is $F=2$.
        }
\label{Fig11}
\end{figure}

Fig. \ref{Fig11} shows the waiting time distribution for the chain length $N=128$, the driving force $F=2$ and different densities of the chain $\phi$. The obvious feature is that the waiting times almost experience the same pathway for smaller $s$, due to lower densities of the chain in the nanocontainer.
% which is similar to the results in Ref. \cite{Luo4}
As observed for translocation into the unconfined system \cite{Luo1,Huopaniemi,Luo4}, we observe almost symmetric distributions with respect to the middle monomer $s=N/2$ for weak confinements such as $\phi=0.1$, due to the effective driving force $F(1-f(\phi)/F) \approx F$.
With increasing $\phi$, the waiting times increase only after a certain $s$, which is a little different compared with previous studies \cite{Luo4} for polymer translocation into a two-dimensional fluidic channel with a narrow channel width where the waiting times always increase. Particularly, it takes much longer time for beads $s > N/2$ to exit the pore. For $\phi \approx 0.35$, the waiting times approximately saturate after $s > N/2$. But for $\phi=0.40$ and 0.45, the waiting times increases very rapidly after $s \approx 100$ because $f(\phi)$ is important and the effective driving forces $F(1-f(\phi)/F)$ are greatly reduced. A big difference towards the end of the packaging is due to less backward motion for several end monomers, which leads to shorter waiting times.

\begin{figure}
\includegraphics*[width=\figurewidth]{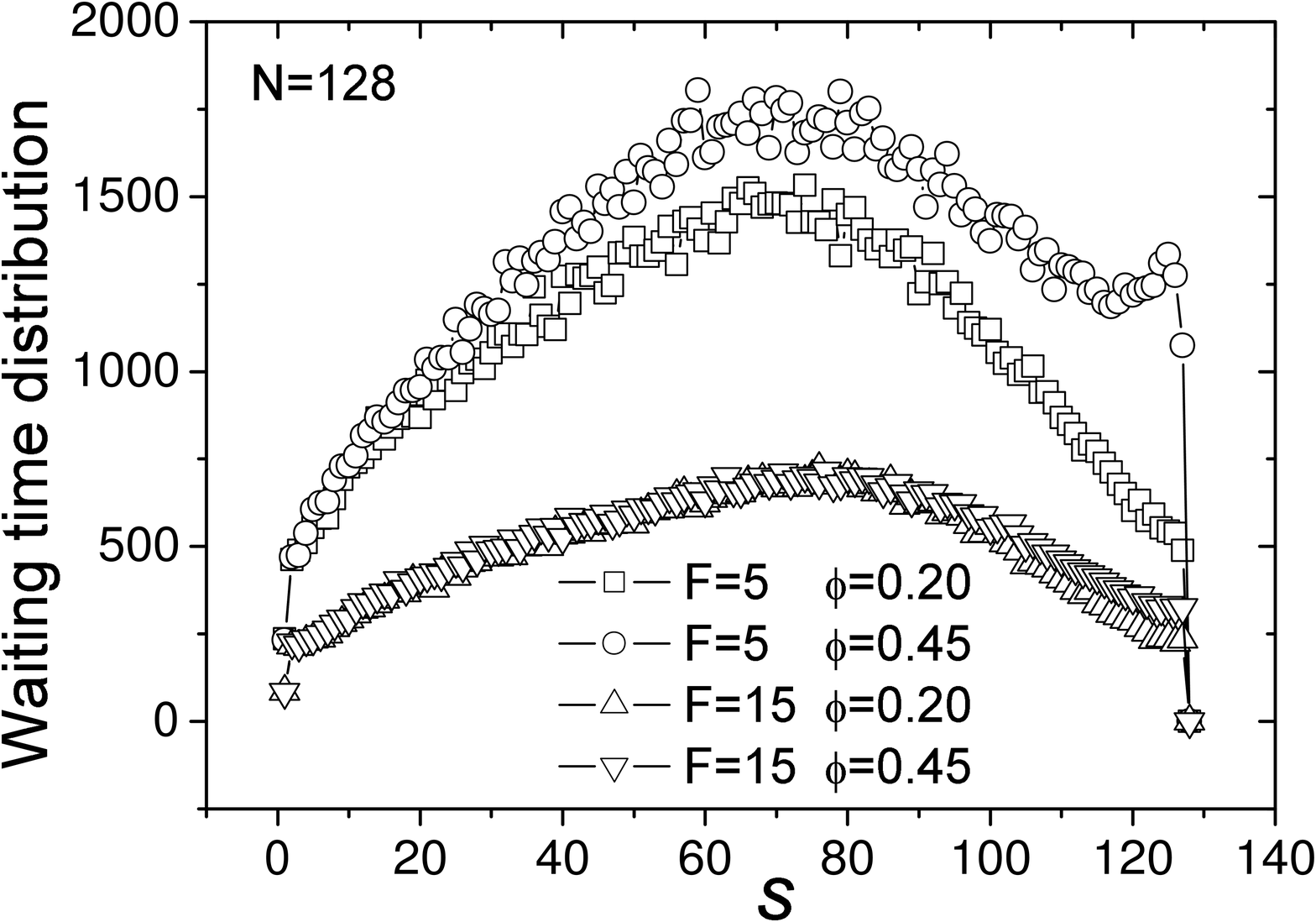}
\caption{Waiting time distribution under different driving forces $F$. The chain length $N = 128$ and density of the chain $\phi =0.2$ and $0.45$.
        }
\label{Fig12}
\end{figure}

For strong driving forces, the effect of $f(\phi)$ on the waiting times is reduced because even if $\phi=0.45$ the effective driving forces $F(1-f(\phi)/F)$ are more and more close to $F$ with increasing $F$, see Fig. \ref{Fig12}. For $F=15$, the waiting times are almost not changed for $\phi=0.2$ and 0.45.

%%%%%%%%%%%%%%%%%%%%%%%%%%%%%%%%%
\section{CONCLUSIONS} \label{chap-conclusions}
%\textit{Conclusions}.

Using Langevin dynamics simulations, we investigate the dynamics of polymer translocation into a circular nanocontainer through a nanopore under a driving force $F$. We observe that the translocation probability initially increases and then saturates with increasing $F$, independent of $\phi$, which is the average density of the whole chain in the nanocontainer. The translocation time distribution undergoes a transition from a Gaussian distribution to an asymmetric distribution with increasing $\phi$. Moreover, we find a nonuniversal scaling exponent of the translocation time as chain length, depending on $\phi$ and $F$. These results are interpreted by the translocated chain conformation in the nanocontainer and the time of an individual segment passing through the pore during translocation.

Our findings have also shed light on dynamics of the packaging of DNA inside virus capsid. Certainly, it is still difficult to directly compare our results with DNA packaging because DNA is a semiflexible polymer which can get entangled and knotted in three dimensions.
In future studies, it would be interesting to consider the influences of the shape of the nanocontainer and the stiffness of the chain on the translocation dynamics.

%%%%%%%%%%%%%%%
\begin{acknowledgments}
This work is supported by the `` Hundred Talents Program '' of CAS and the National Natural Science Foundation of China (Grant Nos. 21074126, 21174140).
\end{acknowledgments}
%%%%%%%%%%%%%%%

\end{document}